\documentclass[acus]{PAC2001}

\usepackage{graphicx}

\begin{document}
\title{THE NUMI HADRONIC HOSE}
\author{R. Ducar,J. Hylen, C. Jensen, M. May, D. Pushka, W. Smart, J. Walton, FNAL, Batavia, IL 60510\\
M. Messier, Harvard University, Cambridge, MA 02138\\
V. Garkusha, F. Novoskoltsev, V. Zarucheisky, IHEP, Protvino, Russia \\
G. Unel, Northwestern University, Evanston, IL 60208 \\
S. Kopp\thanks{kopp@hep.utexas.edu}, M. Kostin, A. Lyukov, R. Zwaska, University of Texas, Austin, TX 78712\\
R. Milburne, Tufts University, Medford, MA  02155}
\maketitle

\vskip .3cm
\begin{abstract}
The Neutrinos at the Main Injector (NuMI) beam supplies an intense $\nu_{\mu}$ beam to the 
Main Injector Neutrino Oscillation Search (MINOS).  The $\nu_{\mu}$'s are derived from
a secondary $\pi^+$ beam that is allowed to decay within a 675~m decay tunnel.  We are
developing a continuous toroidal magnetic focusing system, called the Hadronic Hose,
to better steer this secondary beam.  The Hose will both increase the net 
neutrino flux reaching the MINOS detectors and reduce systematic differences in the
neutrino energy spectra at the two detectors due to solid angle acceptances.

\end{abstract}

\section{NUMI BEAMLINE}

NuMI is a tertiary beam from the Main Injector \cite{NuMI}.  Protons will be 
extracted via single turn extraction ($8.6 \mu$sec pulse, cycle time $1.9$~sec.) 
from the Main Injector and focused downward by 58 mRad, where 
they strike a 0.94~m graphite target to produce the secondary hadron ($\pi$, $K$) beam.  
It is anticipated to achieve $4\times 10^{13}$ protons/pulse.

Two toroidal magnetic "horns" sign- and momentum-select the secondary beam and are 
moveable so as to achieve different neutrino energy spectra\cite{horn}.  
After being focussed forward, the hadrons enter a 675 m long, 
1 m radius evacuated decay volume.  Soft pions tend to decay in the upstream
end of the decay pipe, so must decay at wide angles to send neutrinos to the MINOS detectors, while
stiff pions tend to travel further down the decay pipe.

\begin{figure}[htb]
\centering
\includegraphics*[width=70mm]{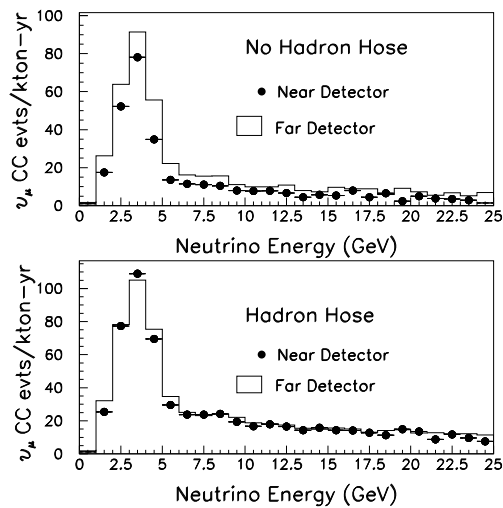}
\caption{Neutrino energy spectra in the far and the near ($\times10^{-6}$) MINOS detectors with and
without the Hadronic Hose for the NuMI low-energy beam (PH2LE).}
\label{lespectra}
\end{figure}

MINOS \cite{MINOS} is a 2-detector neutrino experiment.  The near detector measures the neutrino energy 
spectrum and rate produced at Fermilab.
A second, ``far'', detector is located 735 km away in the Soudan mine in
Minnesota.  The far detector looks for
differences from the near spectrum that could be explained by new physics such as neutrino oscillations.
Figure~\ref{lespectra} shows the expected spectra in the two detectors in the absence of new physics.

\begin{figure}[htb]
\centering
\includegraphics*[width=80mm]{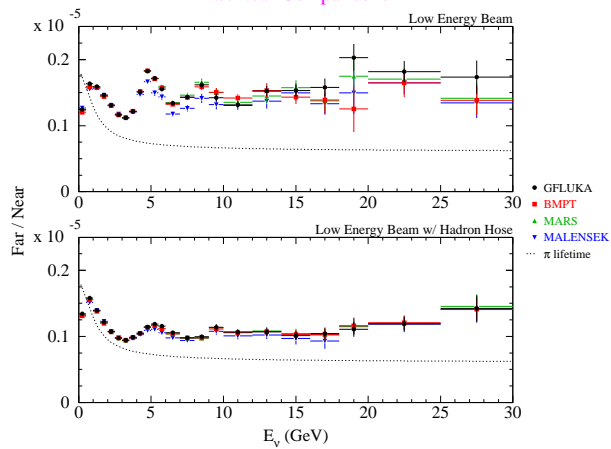}
\caption{``Far over Near Ratio'',F/N,  calculated with Geant/FLUKA, MARS, 
BMPT, and Malensek models, along
with a naive calculation assuming just the pion lifetime dominates this calculation for the PH2LE beam.}
\label{farnear}
\end{figure}

It is desirable to have the spectra at the two detectors as similar as possible.  Differences due to
beamline acceptances, etc., must be characterized by the ``far-over-near ratio'' 
(see Figure~\ref{farnear}), which is the factor by which the near spectrum would be multiplied
to predict the far spectrum in the absence of new physics.  Systematic errors on this 
factor limit the ultimate reach of oscillation searches.

The neutrinos in the peaks of the spectra of Figure~\ref{lespectra} come from pions focused by the
horns, whereas the neutrinos in the high energy tail come from poorly focused pions (pions that
travel through the necks of the horns). 
The high energy tail is nonetheless crucial to the experiment because for low $\Delta m^2$ signals 
it provides us it provides a control sample to demonstrate a region without oscillation effects.

\section{Hadronic Hose}

It is proposed to add an additional focusing system for the secondary pion beam, the
'Hadronic Hose.' It consists of a 2.38~mm diameter aluminum wire at the center of the NuMI decay
tunnel and carries a 1000~A peak current pulse.  The current provides a toroidal
magnetic field which focuses positive particles back toward the center of the NuMI decay pipe.  
A typical meson executes 3-4 orbits around the wire over the length of the decay pipe.
The idea is similar to a proposal by van der Meer, but here implemented for a neutrino beam.\cite{vanderMeer}

The hadronic hose contributes two essential features:  (1)  $\pi$'s and $K$'s that otherwise diverge out to the 
decay pipe walls and interact before they decay are given a restoring force back to the decay pipe center.
Thus, $\pi$'s/$K$'s travel farther and have a greater chance to decay, so the neutrino 
flux is increased (see Table~\ref{fluxes});  (2) the pion orbits effectively randomize
the decay angle between the pion direction and the neutrino that hits the MINOS near and far detectors.
Because of the kinematic correlation $E_{\nu}=0.43E_{\pi}/(1+\gamma^2 \theta^2)$ between 
neutrino energy and decay angle, this randomization is important to make the neutrino energy spectra in 
the near and far MINOS detectors more similar
(see Figure~\ref{lespectra}).  Differences in the far-over-near ratio 
due to different predictions\cite{BMPT,MARS,Malensek,GFLUKA} of pion cross sections off the target
are reduced (see Figure~\ref{farnear}).

\begin{table}[hbt]
\begin{center}
\begin{tabular}{|l|c|c|}
\hline
\textbf{Beam} & \textbf{Flux in peak} & \textbf{Overall Flux} \\ \hline
PH2LE     & 261 & 474 \\
PH2LE-HH  & 327 & 732 \\
PH2ME     & 1077 & 1268 \\
PH2ME-HH  & 1281 & 1675 \\ 
PH2HE     & 2694 & 2745 \\ 
PH2HE-HH  & 2870 & 2983 \\ 
\hline
\end{tabular}
\caption{Charged current event yields in the far MINOS detector in the 3 different
beam configurations, with and without Hadronic Hose.  The peak refers to the regions
$E_{\nu}<$6, 9, and 30 GeV, for low, medium, and high energy beams, while overall refers to $E_{\nu}<40$ GeV. 
The units are events per kiloton of far detector mass per year of running.}
\label{fluxes}
\end{center}
\end{table}

Besides minimizing the experiment's sensitivity to variations in pion productions cross sections,
the Hadronic hose loosens accuracy criteria for other beamline components.  The eccentricity of the
inner conductor of Horn 1, required to be $<$0.08~mm at its neck position in order to 
minimize fringe fields which affect high energy pions, is relaxed to $<$0.12~mm.
The spatial alignment of Horn 1 transverse to the beamline is relaxed from $\pm0.8$~mm to $\pm1.0$~mm.
Relative current variations between the two horns can now be as large as $\pm1.5$\% (c.f. $\pm1.0$\%).  
Finally, the alignment of the straightness of the NuMI decay pipe, previously required to 
be within 9~mm all along its 675~m length, is now relaxed considerably.  
The hose guides the beam along 
even long arc-like excursions in the beamline geometry.

\section{Hose Mechanical Design}

The hadronic hose is built up out of 72 9~m sections of wire, each of which is an independent circuit.
The electrical connection of each section
to the transmission line passes through an electrical feedthrough in the decay pipe wall consisting of
an aluminum pin and ceramic insulating jacket.  The
wire is suspended at the center of the decay pipe by 0.1'' diameter Invar guide wires attached to the decay pipe 
at $1~$m intervals along the decay pipe.  The guide wires are insulated from the decay pipe wall 
by ceramic pins.  The ceramics will
be exposed to $\sim10^9$~Rad/year in NuMI.  Tensioning springs at the end of the wire 
maintain a 2 lb. tension on the wire.  A 2~m hose segment at the most upstream end of the decay pipe is
left unpulsed and absorbs some of the unreacted proton beam.

\begin{figure}[t]
\centering
\includegraphics[width=35mm]{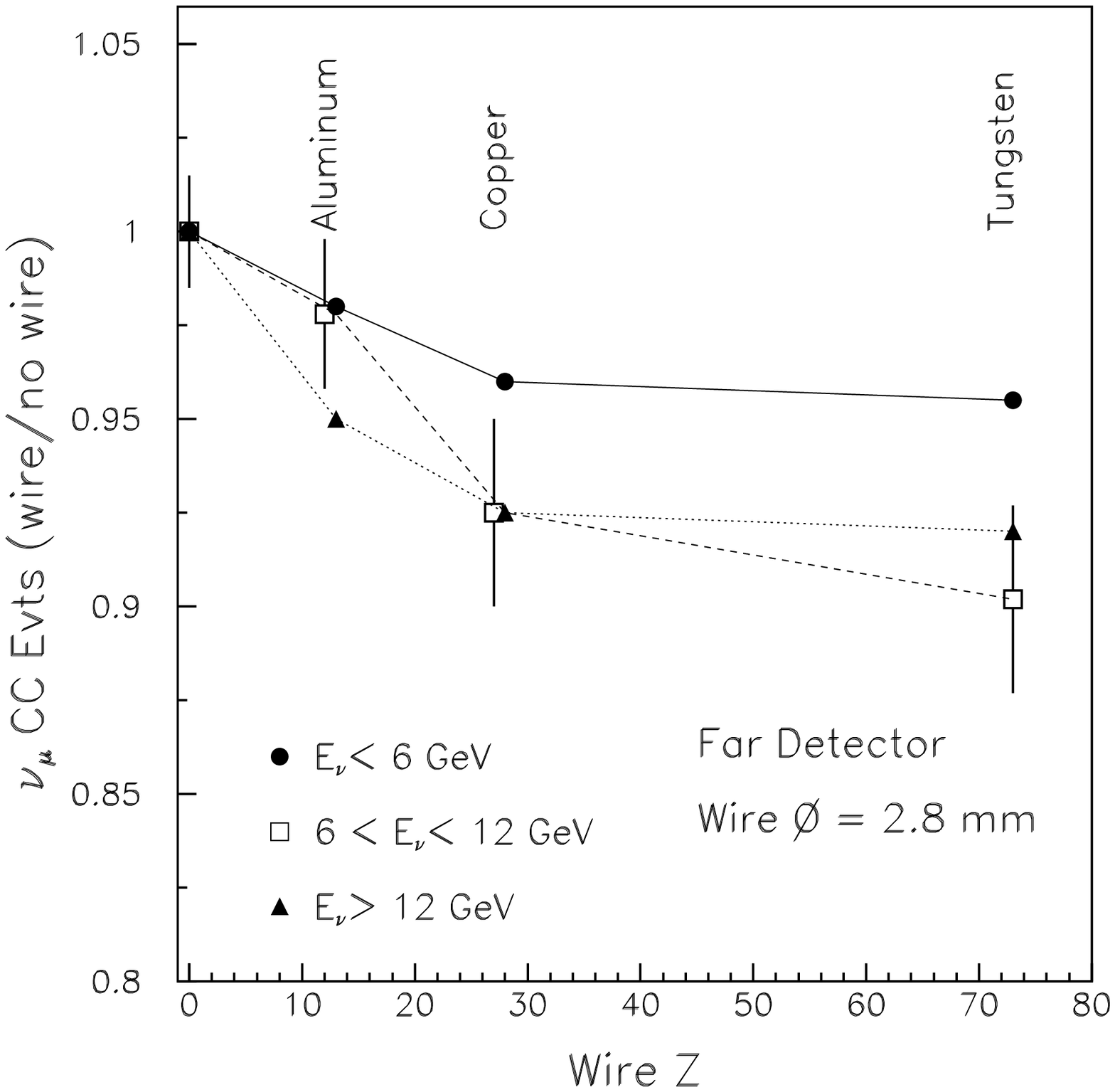}
\includegraphics[width=40mm]{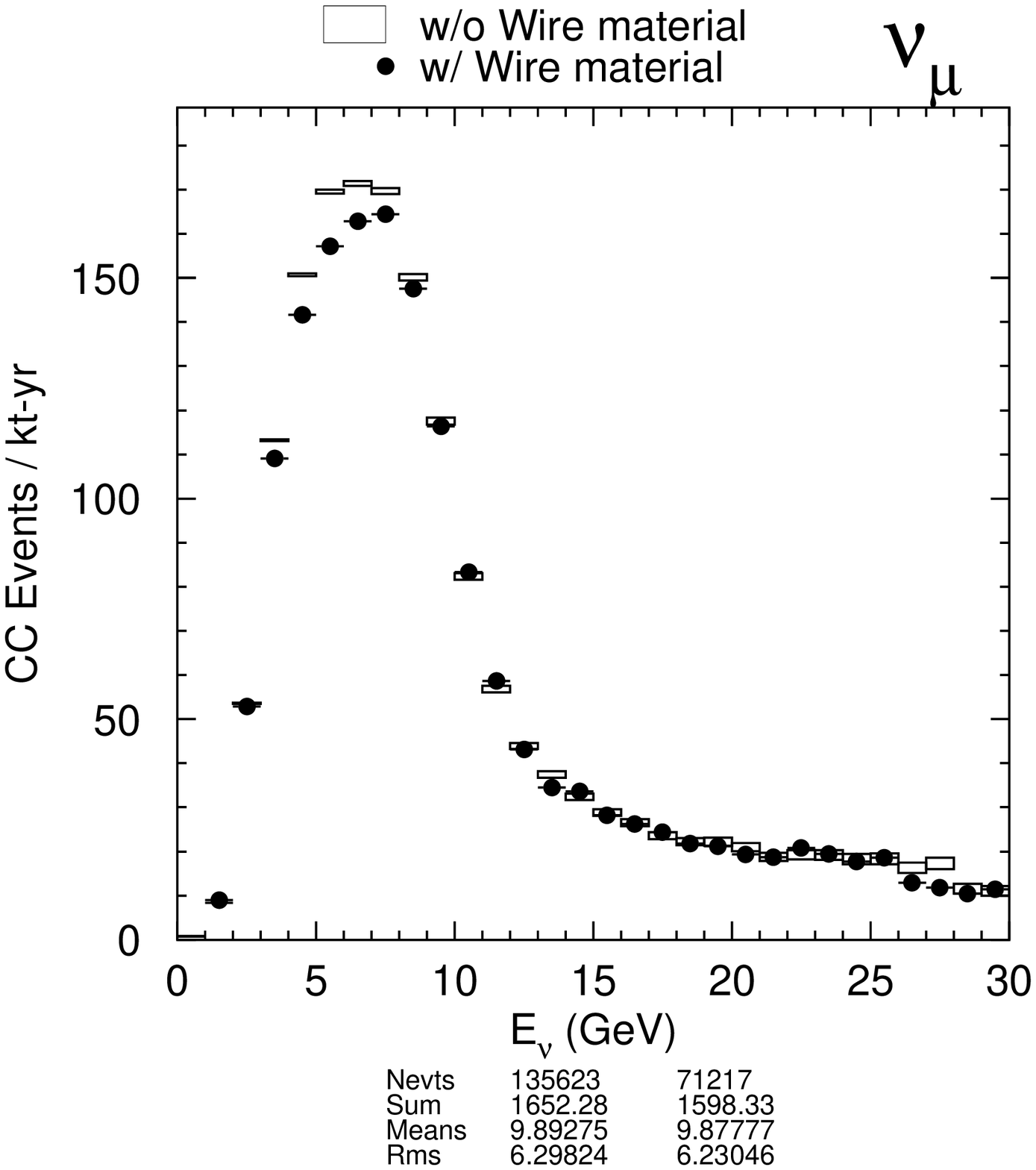}
\caption{Monte Carlo predictions of the 
effect of pion interactions in the wire material on the neutrino flux.  (left)
Dependence on choice of wire meaterial; (right) Monte Carlo spectrum for the
far detector for aluminum wire and wire made of 'vacuum' (PH2ME beam).}
\label{material}
\end{figure}

\section{Hose Electrical Design}

Each hose segment is connected in parallel to a transmission line that runs down the side
passageway of the decay tunnel, through a 3:1 transformer.
The transmission line is energized with a 5000~V pulse 620 $\mu$sec in duration.  The upstream
end of a hose segment is 'center-tapped' to the same transformer as the downstream end of the
preceding hose segment.  All the hose segments are effectively in series, receiving the
same currents to within 1~A according to simulations, where the difference is due to the 80$^o$~C
temperature variation of the hose segments inside the decay tunnel.  If a hose segment should fail
and sever during operation, all the current passes through the transmission line.  

The 620~$\mu$sec RMS, 1000~A peak, pulse causes a 250~V voltage drop across each of the 9~m segments,
dominated by the inductive voltage drop ($L=13\mu H$ for the wire in the 1~m vacuum decay pipe).  The
resistance of the 9~m long, 2.4~mm diameter Al1350 wire is $\sim~83$~m$\Omega$, giving a resistive 
drop of 83~V.  This current pulse deposits $\int i^2 r dt~=~30$~J into the wire.  At 250~V, the voltage
on the wire is well below the breakdown point measured for gas pressures of 0.1-1.0~Torr expected
in NuMI (see Figure~\ref{heat}).

\section{Hardware R\&D}

\begin{figure}[b]
\centering
\includegraphics[width=40mm]{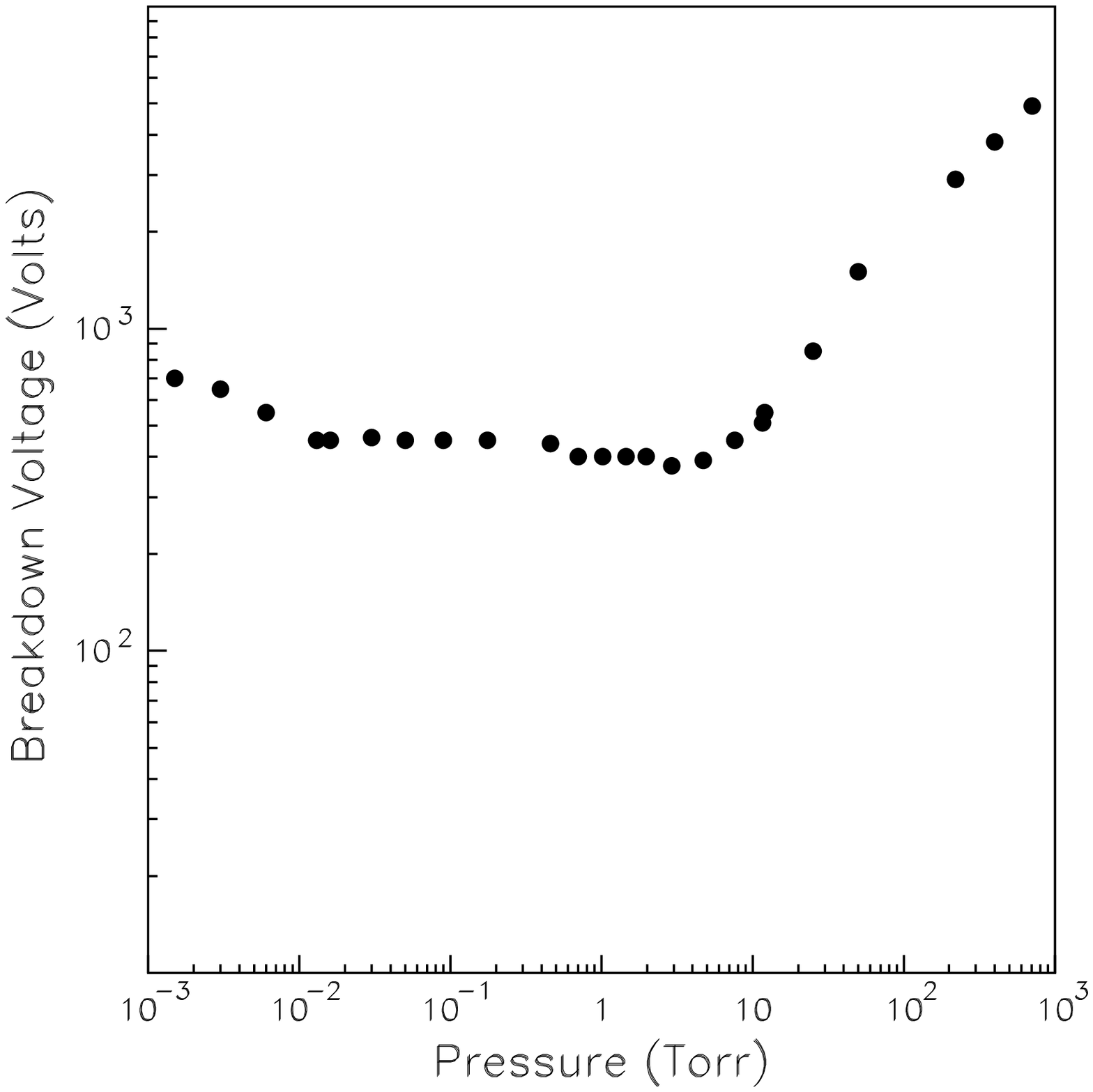}
\includegraphics[width=37mm]{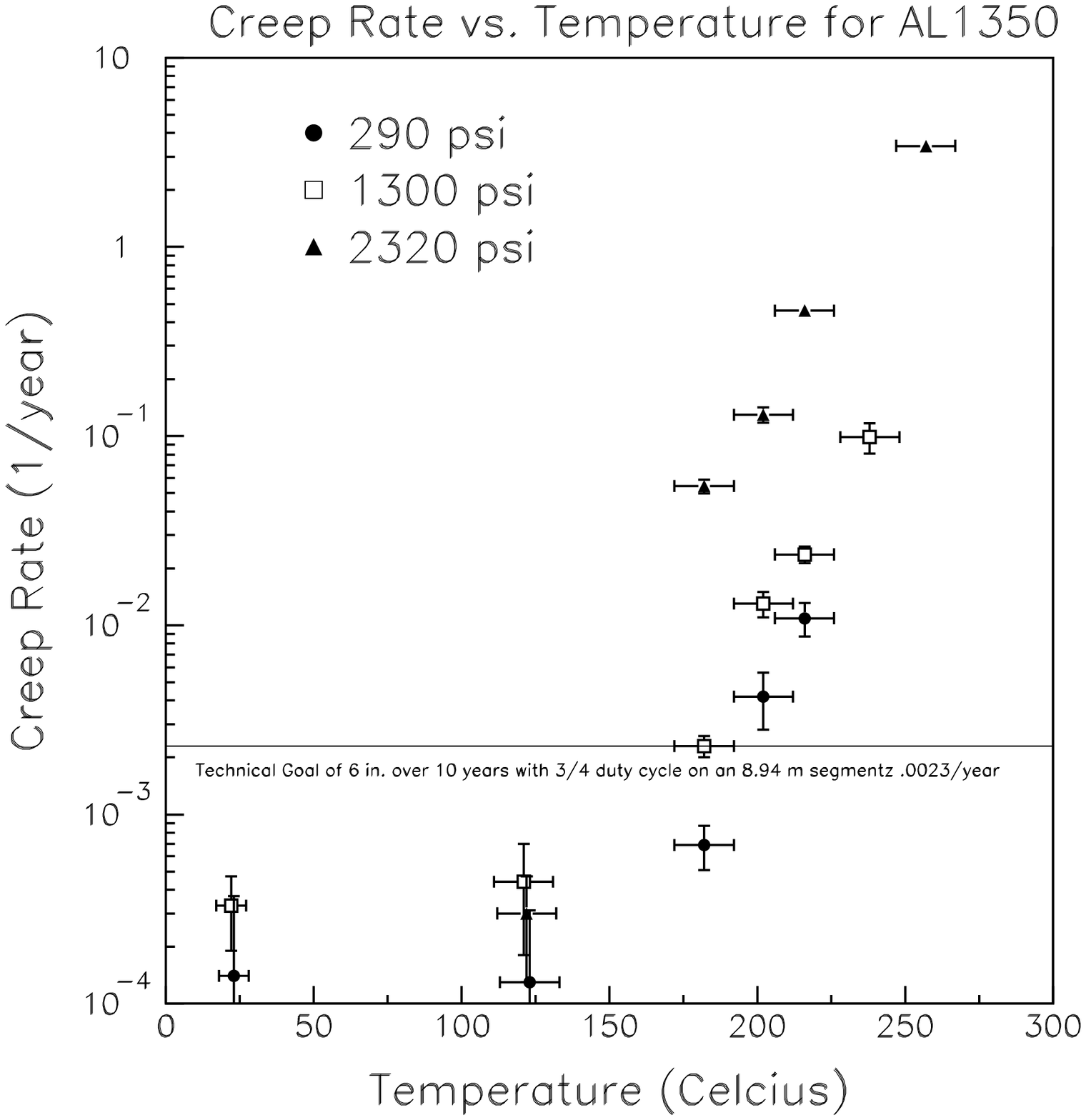}
\caption{(left) Measured breakdown voltage of an aluminum wire in a 6'' diameter vacuum chamber as a 
function of chamber pressure.  (right) Measured creep rate of Aluminum alloy 1350 (anodized) as a 
function of temperature and tension on the wire.}
\label{heat}
\end{figure}

Pion interactions in the wire material motivate the choice of low $Z$ wire (see Figure~\ref{material}),
and also small (2.4 mm) diameter wire to avoid reductions in neutrino flux.
Aluminum alloy 1350 has a conductivity approximately 80\% of pure copper.  Pure
aluminum experiences plastic flow (``creep'') at elevated temperature, and we have measured several 
alloys' creep rates at various temperatures and tensions on the wires over a period of 400~days
by placing them in a high temperature vessel (see Figure~\ref{heat}).  In the NuMI beam it is expected the
wire will operate at 120 - 150 $^o$C, at which the creep rate is low enough to allow expansion into 
the 20~cm gaps between wire segments over 10~years of operation without causing voltage breakdown between
segments.

\begin{figure}[htb]
\centering
\includegraphics*[width=70mm]{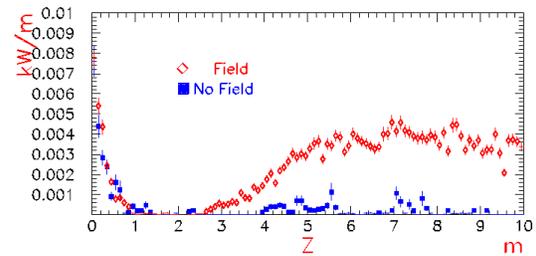}
\caption{Energy deposited in the Hadronic Hose wire from unreacted protons, 
for the hose current on and off, as a 
function of length along the hose.  The first 2 m of hose wire is not 
pulsed and absorbs 'head-on' protons. } \label{deposition}
\end{figure}

The hose wire is hit by the protons that didn't react in the NuMI target.  Approximately
$5\times10^{12}$ protons per beam pulse enter the decay pipe with a $\approx~2$~cm spot size
after multiple scattering in the target, and this cloud
is almost perfectly focused into the hose wire.\footnote{It is an interesting point that the secondary
pion beam does not all interact in the hose wire as well.  In contrast to the proton beam, the
pions multiple scatter through the horn material (20\%$X_0$), so acquire a component of their momentum
in the 'azimuthal' direction looking down the beamline which causes their orbits in the hose field
circle the wire.}  The energy deposition in the hose
wire, as calculated using the MARS\cite{MARS} beamline Monte Carlo, agrees qualitatively with the 
energy deposition expected from protons
focused back into the hose further downstream in the decay tunnel and
from the proton beam directly hitting the end of the first hose wire segment (see Figure~\ref{deposition}).  
The peak of this energy deposition is 18~m downstream in the decay pipe, and is 4.5~W/m.

The hose wire dissipates heat largely through 
conduction through the residual gas ($P=1~$Torr in the decay
pipe) or through blackbody radiation.  The hose wire is coated with a 17~$\mu$m thick aluminum oxide layer
\footnote{This coating also reduces the creep rate by $\sim$10\%.} 
to improve its emissivity from 0.1 for bare aluminum to $\varepsilon \sim 0.6$.  The emissivity
has been measured in a 25 foot long, 6'' diameter vacuum chamber with blackened interior walls and a 
hose wire running down its center at DC current to provide a known power in.   
The temperature was inferred from the elongation
of the wire viewed through a window and the coefficient of thermal expansion measured for the wire.

\section{Conclusions}

We have investigated the potential impact of a new focusing system, the Hadronic Hose, for conventional
neutrino beams.  Such a system is being considered for the NuMI beam at Fermilab, and may be of benefit
to future conventional 'super beams' because of its increase in neutrino flux and ability to control 
systematic uncertainties due to particle production in the target or imperfections in the rest of the 
neutrino beam.


\begin{thebibliography}{9}   



\bibitem{NuMI}  J. Hylen {\it et al.}  ``Conceptual Design for the Technical Components
   of the Neutrino Beam for the Main Injector (NuMI),'' Fermilab-TM-2018, Sept., 1997.
\bibitem{horn} V. Garkusha, F. Novoskoltsev, and V. Zarucheisky, ``The PH2M Two Horn 
   Focussing System for the NuMI Project,'' Fermilab note NuMI-B-471 (1999).
\bibitem{MINOS}  The MINOS Collaboration, ``The MINOS Detectors Technical Design Report,''
   Fermilab NuMI-L-337, Oct. 1998, S. Wojicki, spokesman.
\bibitem{vanderMeer} S. van der Meer, {\it The Beam Guide}, CERN preprint number CERN-62-16, April 17, 1962.
\bibitem{MARS} N.V. Mokhov, ``The MARS Monte Carlo'', Fermilab FN-628 (1995);
   O.E. Krivosheev {\it et al.}, Proc. of the Third and Fourth Workshops on 
   Simulating Accelerator Radiation Environments (SARE3 and SARE4), Fermilab-Conf-98/043(1998) and
   Fermilab-Conf-98/379(1998). 
\bibitem{BMPT} M. Bonesini, A. Marchionni, F. Pietropaolo, and T. Tabarelli de Fatis,
   ``On Particle Production for High Energy Neutrino Beams,'' Eur. Phys. J. {\bf C20}, 13-27 (2001).
\bibitem{Malensek} A.J. Malensek, ``Empirical Formula for Thick 
   Target Particle Production,'' Fermilab FN-341, Oct. 1981.
\bibitem{GFLUKA} {\it GEANT Detector Description and Simulation Tool},
   CERN Program Library, W5013 (1994).
\bibitem{hose} J. Hylen {\it et al.}, ``Proposal to Include Hadronic Hose in the NuMI
   Beamline,'' Fermilab NuMI-B-542, Oct. 1999.
\bibitem{tdr} J. Hylen {\it et al.}, ``Technical Design Report for the Hadronic Hose'',
   NuMI-B-610 (2000).

\end{thebibliography}
\end{document}